\documentclass[hyper]{JHEP} 

\usepackage{epsfig}

\newcommand\fverb{\setbox\pippobox=\hbox\bgroup\verb}

\newcommand\fverbdo{\egroup\medskip\noindent%

            \fbox{\unhbox\pippobox}\ }

\newcommand\fverbit{\egroup\item[\fbox{\unhbox\pippobox}]}

\newbox\pippobox

\title{Anti-de Sitter D-brane dynamics}

\author{J. Kluso\v{n}\\

Department of
Theoretical Physics and Astrophysics\\
Faculty of Science, Masaryk University\\
Kotl\'{a}\v{r}sk\'{a} 2, 611 37, Brno\\
Czech Republic\\

    E-mail: \email{klu@physics.muni.cz}}
\author{Kamal L. Panigrahi\footnote{Address after 1 October 2005: ``Dipartimento di Fisica (DIFI),
Universita di Genova, Via Dodecaneso 33, I-16146 Genova, Italy'' }\\

Dipartimento di Fisica,\\
Universita \& I.N.F.N. Sezione di Roma 2, ``Tor Vergata'' \\
Via della Ricerca Scientifica 1, 00133  Roma   ITALY\\

    E-mail: \email{Kamal.Panigrahi@roma2.infn.it}}

\preprint{ROM2F/2005/20 \\
\hepth{0509065}}

\abstract{We study the anti-de Sitter D-brane dynamics in
Hamiltonian formulation. By exploiting the existence of certain
conserved charges, we solve the equation of motion for a D1-brane
in the AdS$_3$ background and find the space-time dependent
solutions. We further study the dynamics by mapping the problem to
that of the open string tachyon, and examine the time dependent
solutions in some detail. } 


\def\pb #1{\left\{#1\right\}}

\def\mT{\mathcal{T}}

\def\bA{\mathbf{A}}
\def\mE{\mathcal{E}}
\def\mK{\mathcal{K}}
\def\mH{\mathcal{H}}
\def\bAi{\left(\mathbf{A}^{-1}\right)}
\def\mL{\mathcal{L}}
\begin{document}
\section{Introduction}\label{first}
Study of string theory in curved backgrounds is of great interest.
Among these, Anti-de Sitter (AdS) spacetime has played a very
important role in the evolution of string theory. The famous
holographic AdS/CFT correspondence \cite{Maldacena:1997re} relates
quantum gravity in AdS space to a conformal field theory at the
boundary. A concrete realization of this conjectured duality comes
from the study of string theory on the SL$(2,\mathbf{R})$
Wess-Zumino-Witten model and its discrete orbifolds. The
corresponding target space geometries are in general three
dimensional AdS space with NS-NS three form flux. These
backgrounds are considered to be the laboratory for testing the
AdS/CFT duality beyond the supergravity approximation \cite{AdS3}.
D-branes in these backgrounds are of interest and has been studied
extensively in \cite{Bachas:2000fr,D-AdS3}.

The time dependent physics in string theory is equally challenging
that includes many puzzling issues. Recently, a classical time
dependent solution of the open string theory was suggested in
\cite{Sen:2002nu} which describes the rolling of the tachyon in
the valley of its potential on a unstable D-brane (or in
brane-anti brane pair). The late time product of this 'rolling'
has been interpreted as a classical 'tachyon
matter'\cite{Sen:2002in} state that is devoid of any obvious open
string excitations, and which has the properties of the
non-rotating, non-interacting, pressureless dust\cite{Sen:2002an}.
An important outcome of the recent observation of the tachyon
condensation is that an effective action of the Dirac-Born-Infeld
(DBI) type essentially captures all the intriguing aspects of the
rolling tachyon solution of the full string theory. See
\cite{Sen:2004nf} for a detailed review and a comprehensive list
of references on the open string tachyon dynamics. More recently
Kutasov \cite{Kutasov-RB,Kutasov-GT} gave a geometrical
interpretation of the open string tachyon in the form of a rolling
brane--the time dependent dynamics of the D-brane  in the vicinity
of a stack of NS5-branes, by using the DBI analysis. It was
extended further into the full fledged string theory by
constructing the relevant boundary states \cite{NST-BS}. For more
work along this direction see \cite{Rbrane}. However most of the
work has concentrated on the BPS D-brane falling into the
NS5-brane and by showing the similarity between the radial mode
with the perturbative tachyon on the unstable D$p$-brane. But a
similar study of the AdS D-brane dynamics is still lacking. In a
recent paper \cite{Huang:2005hy} some properties of AdS D-branes
has been outlined by using the DBI action on them. Hence a natural
question arises whether one can study the evolution or the rolling
of D-branes in AdS backgrounds by using some kind of 'tachyon
mapping'. We address this problem along with finding out the
solution to the equation of motion of the AdS D-branes in the
hamiltonian formulation. The motivation for studying the present
problem is manifold. Firstly, we extend the radion rolling
dynamics beyond the flat space D-branes. We do this by using the
hamiltonian formulation and by showing the existence of certain
conserved charges. Second, we try to find out the solutions to the
equations of motion for the AdS D-branes that can be interpreted
as the D-brane moving towards the horizon of AdS spacetime.

The plan of the present paper is as follows. In section-2, we
spell out the construction of the DBI action for a BPS D-brane in
a general background in the presence of various worldvolume flux.
In section-3 we focus our attention to the AdS D1-brane moving in
the AdS$_3$ background towards the horizon in a controlled manner.
By showing the invariance of the action under the existence of a
particular symmetry (possibly broken at the Lagrangian level), we
are able to show two different conserved charges--corresponding to
the total energy and the charge corresponding to the conserved
dilatation. We further write down an expression for the conserved
hamiltonian and find out the equations of motion for the D1-brane.
Then we use the tachyon mapping to show the equivalence of the
radial coordinate to that of rolling tachyon of the open string
models, and in particular show that there are no plane wave
solution for the geometrical tachyon near the horizon of the
AdS$_3$ space. This is exactly the same behavior that the tachyon
matter obey at the minimum of the potential, when the D-brane
decays into the vacuum. We further map the solution to the global
coordinate system of AdS$_3$ to study some general properties of
the solution. In section-4, we present our conclusion.

\section{Dirac-Born-Infeld action}
We start with the Dirac-Born-Infeld action for Dp-brane in general
background
\begin{eqnarray}\label{actgen}
S=-\tau_p\int d^{p+1}\xi
e^{-\Phi}\sqrt{-\det \bA_{\mu\nu}} \ ,
\bA_{\mu\nu}=\gamma_{\mu\nu}+F_{\mu\nu} \ ,
\nonumber \\
\end{eqnarray}
where $\tau_p$ is D$p$-brane tension,  $\Phi(X)$ is the dilaton.
The $\gamma_{\mu\nu} \ , \mu,\nu=0,\dots, p$ is embedding of the
metric into the D$p$-brane worldvolume
\begin{equation}
\gamma_{\mu\nu}=g_{MN}\partial_\mu X^M
\partial_\nu X^N \ , \ M,N=0,\dots, D \ .
\end{equation}
In (\ref{actgen}) the induced two-form $F_{\mu\nu}$ is given by
\begin{equation}
F_{\mu\nu}=b_{MN}\partial_\mu X^M
\partial_\nu X^N+\partial_\mu A_\nu -
\partial_\nu A_\mu \ .
\end{equation}
Thanks to the diffeomorphism invariance of the worldvolume theory,
it is natural to fix some of the spacetime coordinates $X^M$ to be
equal to the worldvolume coordinates $\xi^\mu$:
\begin{equation}
X^\mu=\xi^{\mu} \ , \ \mu=0,\dots, p \ .
\end{equation}
Note that this choice generally leads to the induced metric
$\gamma_{\mu\nu}$ and two form $F_{\mu\nu}$ to  be functions of
$\xi $ and $X^I$ where $I,J,...=p+1,\dots, D$ label coordinates
that are transverse to the worldvolume of Dp-brane:
\begin{eqnarray}
\gamma_{\mu\nu}&=& g_{\mu\nu}(\xi, X) +g_{\mu
I}(\xi,X)\partial_\nu X^I+ g_{J\nu}(\xi,X)\partial_\mu X^J
+g_{IJ}(\xi,X)\partial_\mu X^I
\partial_\nu X^J \ , \cr & \cr
F_{\mu\nu}&=& b_{\mu\nu}(\xi, X) +b_{\mu I}(\xi,X)\partial_\nu
X^I+ b_{J\nu}(\xi,X)\partial_\mu X^J +b_{IJ}(\xi,X)\partial_\mu
X^I \partial_\nu X^J \cr & \cr &+&
\partial_\mu A_\nu-\partial_\mu A_\nu
\ .\nonumber \\
\end{eqnarray}
In the following we will consider the case where the metric is
diagonal so that $g_{\mu I}=g_{J\nu}=0$. Now the equation of
motion for $X^I$ derived from \ref{actgen} takes the form
\begin{eqnarray}
&&\frac{\delta e^{-\Phi}}{\delta X^I}\sqrt{-\det\bA}+
\frac{e^{-\Phi}}{2}\left[ \frac{\delta g_{\mu\nu}}{\delta X^I}+
\frac{\delta g_{KL}}{\delta X^I}\partial_\mu X^K
\partial_\nu X^L+\frac{\delta B_{\mu\nu}}{\delta X^I}
\right] \bAi^{\nu\mu}\sqrt{-\det\bA} \cr & \cr
&-&\frac{1}{2}\partial_\mu\left[e^{-\Phi} g_{IK}\partial_\nu
X^K\left(\bAi^{\nu\mu}+\bAi^{\mu\nu}\right)
\sqrt{-\det\bA}\right]\cr & \cr &-& \frac{1}{2}\partial_\mu\left[
e^{-\Phi}
b_{I\nu}\left(\bAi^{\nu\mu}-\bAi^{\mu\nu}\right)\sqrt{-\det\bA}
\right] \cr & \cr &-& \frac{1}{2}\partial_\mu\left[ e^{-\Phi}
b_{IK}\left(\bAi^{\nu\mu}-\bAi^{\mu\nu}\right)\sqrt{-\det\bA}
\right]=0 \ .
\nonumber \\
\end{eqnarray}
Finally, we should also determine the equation of motion
for the gauge field $A_\mu$:
\begin{equation}
\partial_\nu \left[e^{-\Phi}\left(\bAi^{\mu\nu}-
\bAi^{\nu\mu}\right)\sqrt{-\det\bA}\right]=0
\end{equation}
In what follows we restrict ourselves to the examples when the
induced metric does not depend on the worldvolume coordinates
$\xi^{\mu}$. More precisely, we also presume that the metric and
the two form field depends only on a single coordinate, say ${\rm
R}$ which is the reflection of the rotation symmetry of the
background.

\section{D-brane dynamics in AdS$_3$-background}
AdS$_3$ backgrounds offer a laboratory for testing the
gauge-gravity duality beyond the supergravity approximation. Our
aim in this section is to study the time dependent dynamics of the
AdS D-branes. We do this by applying the most general procedure
depicted in the previous section. We concentrate mostly on the
particular case of D1-brane in $AdS_3$ background. The metric and
the NS-NS two form field of such a background is given by
\begin{equation}
ds^2=-r^2dt^2+r^2dz^2+r^{-2}dr^2 \ ,
b=r^2dt\wedge dz
\end{equation}
In this particular parametrization, the horizon corresponds to
$r\rightarrow 0$. Let us consider the D1-brane that is stretched
in $(t, z)$-plane. In this case the only embedding coordinate is
${\rm R}$ that we presume depends both on time and $z$
\begin{equation}
{\rm R}= {\rm R}(z,t)
\end{equation}
We also presume that the gauge potential has a nonzero component
$A_z$ only.
With this ansatz the matrix $\bA_{\mu \nu}$ introduced
in (\ref{actgen}) takes the form
\begin{equation}
\bA_{\mu\nu}=\left(\begin{array}{cc} -{\rm R}^2+\frac{(\partial_0
{\rm R})^2}{{\rm R}^2} & \frac{\partial_0 {\rm R}
\partial_z {\rm R}}{{\rm R}^2}+
{\rm R}^2+\partial_0 A_z \\
\frac{\partial_0{\rm R}\partial_z{\rm R}}{{\rm R}^2}-{\rm R}^2-
\partial_0 A_z & {\rm R}^2+\frac{
(\partial_z {\rm R})^2}{{\rm R}^2}
\\ \end{array}\right) \
\end{equation}
and hence the action takes the
following form
\begin{equation}\label{acAds1}
S=-\tau_2\int dt dz \sqrt{{\rm R}^4+(\partial_z{\rm R})^2-
(\partial_0{\rm R})^2-({\rm R}^2+\partial_0 A_z)^2} \ .
\end{equation}
Observe that under the transformations
\begin{equation}\label{diltra}
{\rm R}'(t',z')=\lambda {\rm R}(t,z) \ , A_z'(t',z')=\lambda
A(t,z) \ , t'=\lambda^{-1} t \ , z'=\lambda^{-1} z \ ,
\lambda=\mathrm{const}
\end{equation}
the action above is invariant in the following sense
\begin{equation}
\int dt' dz' \mathcal{L}({\rm R}',A'_z,t',z') =\int dt dz
\mathcal{L}({\rm R},A_z,t,z) \ .
\end{equation}
Using this fact, we should be able to
find two conserved charges,
corresponding to the conserved energy
$\mE$ and the charge corresponding to
the conserved dilatation.
\subsection{Conserved charges}
Adopting the standard procedure we get the
following form of the world volume
stress energy tensor $\Theta_{\mu\nu}$
\begin{eqnarray}
\Theta^\mu_\nu &=& -\mL\delta^\mu_\nu-\frac{
\eta^{\mu\kappa}\partial_\kappa {\rm R}\partial_\nu {\rm R}}
{\sqrt{{\rm R}^4+(\partial_z{\rm R})^2- (\partial_0{\rm
R})^2-({\rm R}^2+\partial_0 A_z)^2}}\cr & \cr &+&
\frac{\delta^\mu_0 ({\rm R}^2+\partial_0 A_z)} {\sqrt{{\rm
R}^4+(\partial_z{\rm R})^2-
(\partial_0{\rm R})^2-({\rm R}^2+\partial_0 A_z)^2}}\partial_\nu A_z \ . \nonumber \\
\end{eqnarray}
Note that the trace of the stress energy tensor does not vanish
\begin{equation}
\Theta^\mu_\mu =\frac{\eta^{\mu\nu}\partial_\mu {\rm R}
\partial_\nu {\rm R}
-({\rm R}^2+\partial_0 A_z)\partial_0 A_z} {\sqrt{{\rm
R}^4+(\partial_z{\rm R})^2- (\partial_0{\rm R})^2-({\rm
R}^2+\partial_0 A_z)^2}} \ .
\end{equation}
that implies that the Lagrangian given above does not define
conformal field theory. In spite of the fact we can find the
dilatation like charge that is a generator of the transformations
given in (\ref{diltra}) that for small $\lambda$ can be written as
\begin{equation}
{x'}^\mu=(1+\epsilon)^{-1}x^\mu =
x^\mu-\epsilon x^\mu
\end{equation}
\begin{equation}
{\rm R}'(\mu')=(1+\epsilon){\rm
R}(x^\mu)= {\rm R}(x^\mu)+\epsilon {\rm
R}(x^\mu) \ ,
\end{equation}
\begin{equation}
A_z'(\mu')=(1+\epsilon)A_z(x^\mu) =
A_z(x^\mu)+ \epsilon A_z(x^\mu) \
\end{equation}
and hence we obtain following form of
the dilatation current $j^\mu_D$:
\begin{eqnarray}\label{jd}
j^\mu_D &=& \left(-\mL \delta^\mu_\nu +\frac{\delta \mL} {\delta
\partial_\mu {\rm R}}
\partial_\nu {\rm R}+
\frac{\delta
\mL}{\delta \partial_\mu A_z}
\partial_\nu A_z\right)(-x^\nu) -\frac{\delta \mL}{\delta \partial_\mu {\rm R}} {\rm R}
-\frac{\delta \mL} {\delta \partial_\mu A_z}A_z \ , \cr & \cr
j^0_D(z)&=& - t\mH(z)-z\left( \Pi_{\rm R}(z)\partial_z {\rm R}(z)+
\Pi_A(z)\partial_z A_z(z)\right) -\Pi_{\rm R}(z){\rm
R}-\Pi_A(z)A_z \ ,
\nonumber \\
\end{eqnarray}
where $\mH$ is Hamiltonian density and where $\Pi_R(z) \ ,
\Pi_A(z)$ are the momenta conjugate to $R$ and $A_z$:
\begin{equation}
\Pi_R(z)=\frac{\delta \mL}{
\delta \partial_0 R(z)} \ ,
\Pi_A(z)=\frac{\delta \mL}
{\delta \partial_0 A_z(z)} \ .
\end{equation}
Note that $j_D^\mu\neq x^\nu \theta_{\nu}^\mu$ that again shows
that this two dimensional field theory is not the conformal one.

From (\ref{jd}) we obtain the following conserved charge
\begin{equation}\label{Qd}
Q_D(t)=\int_{-\infty}^{\infty}dz
j^0_D(z)=-tH-\int_{-\infty}^{\infty} dz \left(z\left[ \Pi_{\rm
R}\partial_z {\rm R} + \Pi_A
\partial_z A_z\right]
+\Pi_{\rm R} {\rm R}+\Pi_A A_z\right) \ .
\end{equation}
As a check, the Poisson bracket of $Q_D$ with $R$ gives
\begin{eqnarray}
\pb{Q_D,{\rm R}(z)}&=& -t\pb{H,{\rm R}(z)}- \int
dz'\left(z'\partial_z' {\rm R} \pb{\Pi_{\rm R}(z'),{\rm R}(z)}+
{\rm R}(z')\pb{\Pi_{\rm R}(z'),{\rm R}(z)}\right)\cr & \cr &=&
-t\partial_0 {\rm R}(z)- z\partial_z {\rm R}(z)-{\rm R}(z) \ ,
\nonumber \\
\end{eqnarray}
using the canonical relations
\begin{equation}\label{pb}
\pb{\Pi_{\rm R}(z),{\rm R}(z')}=\delta(z-z') \ , \ \
\pb{\Pi_A(z),A_z(z')}=\delta(z-z') \
\end{equation}
and the equation of motions
\begin{eqnarray}\label{ceq}
\partial_0 {\rm R}(z)=\pb{H,{\rm R}(z)} \ ,
\partial_0 \Pi_{\rm R}(z) = \pb{H,\Pi_{\rm R}(z)} \ , \nonumber \\
\partial_0 A_z(z)
=\pb{H,A_z(z)} \ ,
\partial_0 \Pi_{\rm A}(z)= \pb{H,\Pi_{{\rm A}}(z)} \ . \nonumber \\
\end{eqnarray}
In the same way we could study the Poisson bracket of $Q_D$ with
$A_z$.

For later purposes we now determine an explicit form of the
Hamiltonian density. To do this let us consider the Lagrangian
density in the following form
\begin{equation}
\mL=-\sqrt{V-\sum_i (f_i
(\partial_0{\Phi}^i)^2
+B_i\partial_0\Phi^i)}\equiv -\triangle \ ,
\end{equation}
where $V$ contain scalar potential for various fields $\Phi^i$ and
also the spatial gradients of these fields, and $f_i$ and $B_i$
are constants. The conjugate momentum $P_i$ to $\Phi^i$ takes the
form
\begin{equation}\label{Pi}
P_i=\frac{2f_i\partial_0\Phi^i+B_i}{2\triangle}
\ , \partial_0\Phi^i=\frac{1}{2f_i}
\left(2P_i\triangle-B_i\right)
\end{equation}
so that the Hamiltonian density takes  the form
\begin{eqnarray}\label{Hden}
 \mH
&=&\sqrt{\left(V+\sum_i\frac{B_i^2}{4f_i}\right)
\left(1+\sum_i\frac{P^2_i}{f_i}\right)} -\sum_i\frac{B_iP_i}{2f_i}
\ .
\nonumber \\
\end{eqnarray}
In case of D1-brane in $AdS_3$ background the  action was given in
(\ref{acAds1}) so that we get
\begin{eqnarray}
V=(\partial_z {\rm R})^2 \ , f_{\rm R}=1 \ , f_{A_z}=1 \ ,
B_{A_z}=2{\rm R}^2 \
\nonumber \\
\end{eqnarray}
and hence the Hamiltonian density (\ref{Hden}) is equal to
\begin{equation}\label{hden}
\mH= \sqrt{((\partial_z {\rm R})^2+{\rm R}^4) (1+\Pi_A^2+\Pi_{\rm
R}^2)} -{\rm R}^2\Pi_A\equiv \sqrt{\mK}-{\rm R}^2\Pi_A \ .
\end{equation}
It is also instructive to calculate the Poisson bracket of $Q_D$
with $H$. This can be easily performed using  the canonical
relations (\ref{pb}) and the equation of motions (\ref{ceq})
\begin{eqnarray}\label{pqh}
\pb{Q_D,H} &=& \partial_0\Big[ \int_{-\infty}^{\infty}dz
 \Big[ z\left(\Pi_{\rm R}(z)\partial_z {\rm R}(z)
+ \Pi_A(z)\partial_z A_z(z)\right) \cr & \cr &+& \Pi_{\rm R}(z)
{\rm R}(z) +\Pi_A(z) A_z(z)\Big]\Big] = \frac{dQ}{dt}+H \ ,
\end{eqnarray}
using
\begin{equation}
\pb{A(z)B(z),H}= \pb{A(z),H}B(z)+A(z)\pb{B(z),H} \
\end{equation}
and also the fact that
\begin{eqnarray}
\frac{dQ_D}{dt}&=& -H +
\partial_0\Big(
\int_{-\infty}^{\infty}dz
 \Big[ z\left(\Pi_{\rm R}(z)\partial_z {\rm R}(z)
+ \Pi_A(z)\partial_z A_z(z)\right) \cr & \cr &+& \Pi_{\rm R}(z)
{\rm R}(z) +\Pi_A(z) A_z(z)\Big]\Big) \ .
\end{eqnarray}
Note that (\ref{pqh}) is in agreement with the definition of the
time evolution of $Q_D$
\begin{equation}
\frac{dQ_D}{dt}= \frac{\partial Q_D} {\partial t}+\pb{H,Q_D} \ .
\end{equation}
Let us now study in more details the canonical equation of motion.
Firstly, since the Hamiltonian does not explicitly depend on $A_z$
it follows that $\Pi_A$ is conserved
\begin{equation}
\partial_0 \Pi_A(z) = \pb{H,\Pi_A(z)}=0 \ .
\end{equation}
We can consistently solve the equation of motion with the ansatz
that $\Pi_A=\mathrm{const.}$ with the physical interpretation that
$\Pi_A$ gives the density of fundamental strings.

Using (\ref{hden}) the equation of motion for $\Pi_{\rm R}$ and
${\rm R}$ are equal to
\begin{eqnarray}\label{eqce}
\partial_0 {\rm R}(z)&=&
\frac{\delta H}{\delta \Pi_{\rm R}(z)}= \frac{((\partial_z {\rm
R})^2+{\rm R}^4 )\Pi_{\rm R}}{\sqrt{\mK}} \ , \cr & \cr
\partial_0 \Pi_{\rm R}(z)&=&
-\frac{\delta H}{\delta {\rm R}(z)}= -\frac{2{\rm R}^3(1+\Pi_A^2+
\Pi_{\rm R}^2)}{ \sqrt{\mK}}+2\Pi_A{\rm R} +\partial_z \left[
\frac{\partial_z {\rm R}( 1+\Pi_A^2+\Pi_{\rm R}^2)}{
\sqrt{\mK}}\right] \ , \cr & \cr
\partial_0 A_z(z)&=& \frac{\delta H}{
\delta \Pi_A}= \frac{(\partial_z {\rm R})^2+{\rm
R}^4)\Pi_A}{\sqrt{\mK}}-{\rm R}^2
\ . \nonumber \\
\end{eqnarray}
Our goal is to find the solution of the canonical equations of
motion that describes the time evolution of an inhomogenous
D1-brane. The structure of the Hamiltonian density (\ref{hden})
suggests that we should search for the solution of ${\rm R}$ of
the form
\begin{equation}
\partial_z {\rm R}=k_z {\rm R}^2 \ .
\end{equation}
A general solution of ${\rm R}$ can be given by (with an
integration constant $f(t)$)
\begin{equation}\label{rtz}
\frac{1}{\rm R(z,t)}=- (k_z z + f(t)) \ ,
\end{equation}
and we also get
\begin{equation}\label{hkr}
\mH={\rm R}^2\left( \sqrt{(1+k^2_z)(1+\Pi_A^2+\Pi_{\rm
R}^2)}-\Pi_A\right) \ .
\end{equation}
Using (\ref{hkr}) we express $\Pi_R$ as function of $\mH$ and
${\rm R}$
\begin{equation}\label{pir}
\Pi_{\rm R} = \pm\frac{1}{{\rm R}^2\sqrt{1+k^2_z}}
\sqrt{(\mH+\Pi_A{\rm R}^2)^2-{\rm R}^4(1+k^2_z)(1+\Pi_A^2)} \
\end{equation}
and insert it to the first equation in (\ref{eqce})
\begin{equation}\label{eqh}
\partial_0 {\rm R}
=\pm \sqrt{1+k^2_z}{\rm R}^2 \sqrt{1-\frac{{\rm
R}^4(1+\Pi_A^2)(1+k^2_z)} {(\mH+\Pi_A{\rm R}^2)^2}} \ .
\end{equation}
If we also use the fact that
\begin{equation}
\partial_0 {\rm R}=-\dot{f}{\rm R}^2 \
\end{equation}
the equation (\ref{eqh}) simplifies as
\begin{equation}\label{dotf1}
\dot{f}=\mp \sqrt{1+k^2_z} \sqrt{1-\frac{{\rm
R}^4(1+\Pi_A^2)(1+k^2_z)} {(\mH+\Pi_A{\rm R}^2)^2}} \ .
\end{equation}
Let us insert (\ref{hkr}) into the equation above and we get
\begin{equation}\label{dotf}
\dot{f}=\mp \sqrt{1+k^2_z} \sqrt{1-\frac{1+\Pi^2_A}{
(1+\Pi_A^2+\Pi_{\rm R}^2)}} \ .
\end{equation}
By comparing the left and the right side of this equation  we
deduce that $\Pi_R$ depends on $t$ only. Then the equation of
motion for $\Pi_R$ simplifies as
\begin{eqnarray}\label{dpr}
\partial_0\Pi_{\rm R}-\frac{2{\rm R}}{\sqrt{1+k^2_z}}
\sqrt{1+\Pi^2_A+\Pi^2_R}+2\Pi_A{\rm R} \ .
\nonumber \\
\end{eqnarray}
Moreover, looking at the form of the time and spatial dependence
of ${\rm R}$ given in (\ref{rtz}), one expects that $f(t)$ should
be linear function of $t$ as well. In fact, the worldvolume theory
is still Lorentz invariant. With this presumption $\dot{f}={\rm
const}$, and hence the equation (\ref{dotf}) implies that
$\Pi_{\rm R}={\rm const.}$ as well.  Now the equation (\ref{dpr})
implies
\begin{equation}
\Pi^2_{\rm R}=\Pi^2_A k^2_z - 1
\end{equation}
and from (\ref{dotf}) we get
\begin{equation}
f=\mp t\sqrt{k^2_z-\frac{1}{\Pi^2_A}}+ f_0=k_0 t + f_0 \ , \ f_0 =
{\rm const.}
\end{equation}
where we have introduced the following notation
\begin{equation}
k_0=\mp \sqrt{k^2_z-\frac{1}{\Pi^2_A}} \ , \ \ -k_0^2 +
k_z^2=-\frac{1}{\Pi_A^2} \ .
\end{equation}
With the above, the general solution of ${\rm R}$ in (\ref{rtz})
can be written as
\begin{equation}\label{genrtz}
-\frac{1}{{\rm R}(t, z)} = k_0 t + k_z z + f_0 \ .
\end{equation}
In summary we have found time and space dependent solution of
D1-brane equation of motion using the Hamiltonian formalism. We
will say more words about this solution in the next section, where
we will study the same problem with the help of the Lagrangian
density for the new mode, ``geometrical tachyon'' that we define
below.
\subsection{Tachyon mapping}
In this section, we try to find out the solutions to the equation
of motion for the radial coordinate ('radion') on the D1-brane, by
mapping it to the open string tachyon on the unstable branes. Our
strategy will be as follows. We start with the action for a
D1-brane in the $AdS_3$ background that has already been derived
in the previous section
\begin{equation}\label{acAds}
S=-\tau_2\int d^2x \sqrt{{\rm R}^4+(\partial_1{\rm R})^2-
(\partial_0{\rm R})^2-({\rm R}^2+\partial_0 A_1)^2} \ ,
\end{equation}
where we have replaced $\partial_z {\rm R}\rightarrow\partial_1
{\rm R}, \ {\rm and} \ A_z\rightarrow A_1$ as compared to the
previous section. Now let us map the above action to the tachyon
effective action of the open string model, with the geometric
tachyon $\mT$ being the only dynamical field involved. As opposite
to the standard construction where there is no gauge field
present, here the situation seems to be more complicated. We
propose the following procedure. First we pass to the Hamiltonian
formalism. This was done in the previous section with the
hamiltonian density given by
\begin{equation}\label{hdtm}
\mH= \sqrt{((\partial_1 {\rm R})^2+{\rm R}^4)
(1+\Pi_A^2+\Pi_R^2)}-{\rm R}^2\Pi_A\equiv \sqrt{\mK}-{\rm
R}^2\Pi_A \ .
\end{equation}
As we have shown in the previous section, the conjugate momentum
$\Pi_A$ is a constant. Now we see that the dynamical modes in
(\ref{hdtm}) are $\Pi_R$ and $R$. Then we perform an inverse
Legendre transformation to get the Lagrangian for $R$ that depends
on the constant $\Pi_A$ as well. More precisely, we define the new
Lagrangian density for $R$ as
\begin{equation}
\mL_{\rm{new}}=\Pi_{\rm R}\partial_0{\rm R}-\mH \ .
\end{equation}
Using the fact that
\begin{equation}\label{dorp}
\partial_0{\rm R}= \frac{((\partial_1 {\rm R})^2+{\rm R}^4) \Pi_{\rm R}}{\sqrt{\mK}}
\end{equation}
we get
\begin{equation}
\mL_{\rm{new}} = -\frac{((\partial_1{\rm R})^2+{\rm R}^4)
(1+\Pi_A^2)} {\sqrt{\mK}}+{\rm R}^2\Pi_A \ .
\end{equation}
As a next step, we express $\mK$ that depends on $\Pi_R$ and ${\rm
R}$, as functions of ${\rm R}$ and $\partial_0 {\rm R}$. Using
(\ref{dorp}) we obtain
\begin{equation}
\Pi_{\rm R}= \frac{\partial_0 {\rm R}\sqrt{1+\Pi_A^2}}
{\sqrt{((\partial_1 {\rm R})^2+{\rm R}^4)- (\partial_0{\rm R})^2}}
\end{equation}
and hence
\begin{equation}
\mK = \frac{((\partial_1{\rm R})^2+{\rm R}^4)^2 (1+\Pi_A^2)}
{(\partial_1 {\rm R})^2+{\rm R}^4- (\partial_0 {\rm R})^2} \ .
\end{equation}
Finally we obtain the Lagrangian density $\mL_{\rm{new}}$ in the
form
\begin{eqnarray}\label{ldn}
\mL_{\rm{new}} = -\sqrt{1+\Pi_A^2}{\rm R}^2 \sqrt{1+\frac{1}{{\rm
R}^4} ((\partial_1 {\rm R})^2-
(\partial_0 {\rm R})^2)} +\Pi_A {\rm R}^2 \ .  \nonumber \\
\end{eqnarray}
Using the manifestly covariant Lagrangian density (\ref{ldn})  we
can introduce the so called 'geometrical tachyon' $(\mT)$ in the
following form
\begin{equation}\label{ident}
\frac{d{\rm R}}{{\rm R}^2}=d\mT \ , \ \ \ \mT=-\frac{1}{{\rm R}}.
\end{equation}
Now
\begin{equation}
{\rm R} \rightarrow 0 \Longrightarrow \mT \rightarrow -\infty, \ \
\ {\rm R} \rightarrow \infty \Longrightarrow \mT \rightarrow 0.
\end{equation}
With the identification (\ref{ident}), the Lagrangian (\ref{ldn})
takes the form
\begin{equation}\label{ldnt}
\mL=-\sqrt{1+\Pi_A^2} \frac{1}{\mT^2} \sqrt{1+\eta^{\mu\nu}
\partial_\mu \mT \partial_\nu \mT} + \Pi_A \frac{1}{\mT^2} \ .
\end{equation}
Our goal now is to solve the equation of motion for $\mT$ that is
derived from (\ref{ldnt})
\begin{eqnarray}\label{eqgt}
\frac{2\sqrt{1+\Pi_A^2}} {\mT^3} \frac{1}{\sqrt{1+\eta^{\mu\nu}
\partial_\mu \mT
\partial_\nu \mT}}
+\frac{\sqrt{1+\Pi_A^2}} {\mT^2}
\partial_\mu
\left[\frac{\eta^{\mu\nu}
\partial_\nu \mT}{\sqrt{
1+\eta^{\mu\nu}
\partial_\mu \mT\partial_\nu \mT}}
\right]-
\frac{2\Pi_A}{\mT^3}=0 \ .
\nonumber \\
\end{eqnarray}
Let us presume that there exists a solution, to the equation of
motion (\ref{eqgt}), of the following the form
\begin{equation}\label{an}
\eta^{\mu\nu}\partial_\mu \mT
\partial_\nu \mT={\rm K}^2 \ , \ \ \ \
\eta^{\mu\nu} \partial_\mu \partial_\nu \mT=0 \ ,
\end{equation}
with a constant ${\rm K}$ to be determined. For such an ansatz
(\ref{eqgt}) reduces into
\begin{equation}
\frac{1}{\mT^3} \left(\frac{\sqrt{1+\Pi_A^2}} {\sqrt{1+{\rm
K}^2}}-\Pi_A \right)=0.
\end{equation}
The above equation determines the constant ${\rm K}$ in terms of
$\Pi_A$ as
\begin{equation}\label{kpia}
{\rm K}^2=\frac{1}{\Pi_A^2} \ .
\end{equation}
Let us now try to solve the equation (\ref{an}) with a plane wave
ansatz
\begin{equation}\label{pwgt}
\mT = Ae^{k_\mu x^\mu}  \ .
\end{equation}
For (\ref{pwgt}) the equations (\ref{an}) take the form
\begin{equation}
k_0^2-k_1^2=0 \ , -k_0^2+k_1^2={\rm K}^2
\end{equation}
that however implies ${\rm K}=0$. But that is not consistent with
(\ref{kpia}) where we presume a finite $\Pi_A$. This indicates
that for the geometrical tachyon, there exists no plane wave
solution at the horizon of the AdS space. Infact, we can note that
as the D1-brane rolls towards the horizon, there is no physical
open string excitations left, a fact that is consistent with the
observation \cite{Sen:2002an}

Instead of using the plane-wave ansatz (\ref{pwgt}) let us
consider the linear ansatz
\begin{equation}\label{lgt}
\mT=k_\mu x^\mu+B,
\end{equation}
with a constant $B$. Now, the first condition in (\ref{an}) takes
the form
\begin{equation}
\eta^{\mu\nu} k_\mu k_\nu={\rm K}^2 \ ,
\end{equation}
while the second condition $\eta^{\mu\nu}
\partial_\mu\partial_\nu \mT=0$
is satisfied trivially. Now using (\ref{kpia}) we get
\begin{equation}\label{kz}
k_0^2=k_1^2-{\rm K}^2= k_1^2-\frac{1}{\Pi_A^2} \ .
\end{equation}
In terms of the original  variable ${\rm R}$ we then get the
solution
\begin{equation}\label{Rtz}
-\frac{1}{{\rm R}} = k_0 x^0 + k_1 x^1 + b \ , \ \ b =\rm{const} \
.
\end{equation}
We see that this solution coincides exactly with the solution
(\ref{genrtz}) derived in the previous section in the Hamiltonian
formalism.

We can check the consistency of this solution by considering the
case when $k_0=0$. In this case (from \ref{kz}) we get
\begin{equation}
k_1^2 = \frac{1}{\Pi_A^2}
\end{equation}
and hence
\begin{equation}\label{Rsol}
\frac{1}{{\rm R}}= -k_1 x^1 + b \ .
\end{equation}
First of all, the fact that ${\rm R}$ should be positive implies
that we should take $k_1=-\frac{1}{|\Pi_A|}$. Now we determine the
constant $b$ from the requirement that for $x^1\rightarrow x^1_0$,
${\rm R}$ blows up. Hence we get
\begin{equation}
b = - \frac{1}{|\Pi_A|} x^1_0 \ .
\end{equation}
Taking the above ingredients, we obtain the following solution for
$\rm{R}$ that is spatial dependent
\begin{equation}
{\rm R} = \frac{|\Pi_A|}{x^1-x^1_0}
\end{equation}
that coincides with the solution determined in
\cite{Bachas:2000fr}.

Let us return to the time and space dependent solution for ${\rm
R}$. We see from (\ref{Rtz}) that the point where the D-brane is
stretched to ${\rm R} = \infty$ is time dependent with the time
dependence is governed by the equation
\begin{equation}
0 = k_1 x^1+ k_0x^0 + b \ .
\end{equation}
To study the general properties of the solution (\ref{Rtz}), we
map this to the global coordinates of the $AdS_3$. Recall that in
global coordinates $AdS_3$ background metric and the the
three-form field strength takes the form
\begin{equation}\label{bachasads}
ds^2=L^2[-\cosh^2\rho d\tau^2+
d\rho^2+\sinh^2\rho d\phi^2] \
\end{equation}
and
\begin{equation}
H=dB=L^2\sinh(2\rho)d\rho\wedge d\phi
\wedge dt \ ,
B=L^2\sinh^2\rho d\phi\wedge d\tau \ .
\end{equation}
Poincare and global coordinates are related through the relation
\begin{equation}
x^1\pm x^0 = \frac{1}{r}\left( \sinh \rho\sin \phi\pm
\cosh\rho\sin\tau\right)
\end{equation}
 and
\begin{equation}
r=\cosh \rho\cos\tau
+\sinh\rho\cos \phi \ .
\end{equation}
Now it is easy to see that in global coordinates the solution
(\ref{Rtz}) takes the form
\begin{eqnarray}
(k_1+b)\sinh\rho\sin\phi+ (k_0+b)\cosh\rho\sin\tau  = -1 \ .
\nonumber \\
\end{eqnarray}
This has the form of the solution given in \cite{Bachas:2000fr}.
To see this more clearly recall that in the static case $k_0=0$
the upper equation takes the form
\begin{equation}
-\frac{1}{|\Pi_A|}(1+x^1_0)\sinh\rho\sin\phi
-\frac{1}{|\Pi_A|}x^1_0 \cosh\rho\sin\tau  = -1 \ .
\end{equation}
This is in agreement with the result given in \cite{Bachas:2000fr}
where the solution with the boundary condition $x^1_0=0$ was
given.
\subsection{More observation}
Let us now study the embedding equation for given D1-brane. To
find such an equation we should find the relation between the
variables $z,r,t$ and the coordinates $X^0,X^1,X^2,X^3$ that
define $AdS_3$ as embedded hypersurface in the four dimensional
space through the equation
\begin{equation}\label{adsem}
-(X^0)^2+(X^1)^2+(X^2)^2-(X^3)^2=-1 \ .
\end{equation}
The relation between the Poincare  and the Cartesian coordinates
takes the form
\begin{equation}
X^0+X^1=r \ ,
X^2\pm X^3=rw^\pm \ ,
X^0-X^1=\frac{1}{r}
+rw^+w^- \ ,
\end{equation}
where $w^\pm = x^1\pm x^0$. Alternatively
\begin{eqnarray}\label{glc}
X^2&=& r x^1 \ , X^3 = rx^0 \ , \cr & \cr X^0&=& \frac{1}{2}
(r+\frac{1}{r}+r({(x^1)}^2-{(x^0)}^2)) \ , \cr & \cr X^1 &=&
\frac{1}{2}(r-\frac{1}{r} -r({(x^1)}^2-{(x^0)}^2)) \ .
\nonumber \\
\end{eqnarray}
Since we know that the static D1-brane in $AdS_3$ is described by
the solution $r=\frac{|\Pi_A|}{x^1}$, we obtain from the first
equation in (\ref{glc}) that in global coordinates corresponds to
$X^2=|\Pi_A|$. If we insert this relation into (\ref{adsem}) we
get
\begin{equation}\label{ads2em}
(X^0)^2+(X^3)^2-(X^1)^2 = 1 + \Pi_A^2.
\end{equation}
The above defines AdS$_2$ embedded hypersurface in a three
dimensional space.

For general solution (\ref{Rtz}) we can write (for $b=0$)
\begin{equation}
k_1 X^2 + k_0 X^3= r(k_1 x^1 + k_0 x^0)=-1
\end{equation}
Then we get
\begin{equation}
X^2=-\frac{1}{k_1}(1 + k_0 X^3)
\end{equation}
and hence the embedding equation takes the
form
\begin{eqnarray}
(X^0)^2-(X^1)^2+(1-\frac{k_0^2}{k_1^2}) \left(X^3-\frac{k_0}{k_1^2
(1-\frac{k_0^2}{k_1^2})}\right)^2
=1+\frac{1}{k_1^2}+\frac{k_0^2}{k_1^4 (1-\frac{k_0^2}{k_1^2})}
=1+\Pi^2_A
\nonumber \\
\end{eqnarray}
where we have used from eqn. (\ref{kz}) that $\frac{1}{\Pi^2_A} =
k^2_1 -k^2_0$. One can check once again that the above defines
AdS$_2$ D-brane, by comparing with (\ref{ads2em}). More precisely,
if we define
\begin{equation}\label{ads2emm}
\tilde{X}^0=X^0 \ , \tilde{X}^1=X^1 \ , \tilde{X}^3 =
X^3-\frac{k_0}{k_1^2 (1-\frac{k_0^2}{k_1^2})}
\end{equation}
and then perform the following rescaling
\begin{equation}
(1-\frac{k_0^2}{k_1^2})(\tilde{X}^3)^2 \rightarrow (\tilde{X}^3)^2
\ ,
\end{equation} then we see that (\ref{ads2emm}) coincides with
(\ref{adsem}).
\section{Conclusion}
In this paper, we have studied dynamics of the Anti-de Sitter
D-branes, by using the hamiltonian formulation. By exploiting the
existence of conserved charges, we determine the solution that is
both space and time dependent to the D1-brane equation of motion.
We further use the tachyon mapping to relate the radial mode on
the D1-brane to the open string tachyon. We write down explicitly
the Lagrangian and the equation of motion in terms of the new
geometrical tachyon field. By solving the equations of motion, we
have showed the absence of plane-wave solutions. Infact a linear
ansatz does solve the equations of motion for the geometrical
tachyon field, and that is indeed consistent with the solution
predicted by directly solving the hamiltonian equations. This
solution further reduces to the static solution of the AdS-branes
that has been discussed in the literature earlier. We have also
showed that the D1-brane discussed in this paper, is indeed the
AdS$_2$ brane. The analysis done in this paper can easily be
generalized to the higher AdS spaces in order to study the
dynamics of all the AdS D-brane. For example, one can study the
time evolution of the D2-brane in AdS$_3 \times$ S$^3$ background,
and study the possible stable tubular solutions. It is also
interesting to examine the AdS-brane dynamics in other curved
backgrounds like the NS5-brane and the macroscopic fundamental
string backgrounds. It is worthy examining the brane dynamics from
the gauge theory view point, by using the gauge-gravity duality.
Another interesting problem would be to study the brane dynamics
from the full conformal field theory view point, namely by
constructing the relevant boundary states for these branes, and
studying their properties.

\vskip .3in \noindent {\bf Acknowledgement}

The work of JK was supported by the Czech Ministry of Education
under Contract No. MSM 0021622409. The work of KLP was supported
in part by INFN, by the MIUR-COFIN contract 2003-023852, by the EU
contracts MRTN-CT-2004-503369 and MRTN-CT-2004-512194, by the
INTAS contract 03-51-6346, and by the NATO grant PST.CLG.978785.


\end{document}